\documentclass[twocolumn]{revtex4}
\usepackage{amssymb}
\usepackage{latexsym}
\usepackage{epsfig}

\usepackage{amsmath}
\begin{document}
\title{New Tsallis agegraphic Dark Energy in
Horava-Lifshitz cosmology }
\author{M. Abdollahi Zadeh\footnote{
mazkph@gmail.com}}
\address{Not affiliated with any institution, Kazerun, Iran.}

\begin{abstract}
We investigate the new Tsallis agegraphic dark energy (NTADE) scenario in the framework of Horava-Lifshitz cosmology. Considering interacting and non-interacting scenarios of NTADE with dark matter in a spatially non-flat universe, we investigate the cosmological implications of this model in detail. We obtain the differential equation of the evolution of the density parameter, the equation of state parameter and the classical stability of model. Also, we study the behavior of the deceleration parameter and investigate the nature of the statefinder diagnostics and $\omega_D-{\omega_D}^{\prime}$ plane. We find that phantom crossing cannot occur for the state parameter in this scenario and from the plot of the deceleration parameter, we have observed a transition  from decelerating to accelerating phase of the univese. Also, the sign of the square of the sound speed is negative which means unstable behavior at this scenario. The $\omega_D$ and ${\omega_D}^{\prime}$ have negative values which represents the freezing region at here.
\end{abstract}
\maketitle
\section{Introduction}
Twenty years ago, two groups discovered independently that the Universe has entered a phase of accelerated expansion with a redshift between .5-1\cite{Riess,Perlmutter}. Since ordinary matter, such as baryon, has a positive pressure, it cannot accelerate the expansion of the Universe, so it must provide an unusual, unknown substance, called dark energy, the invisible fuel for this accelerated expansion \cite{Sahni2004,Carroll}. The simplest model to solve this problem is to consider vacuum energy as this invisible fuel but, assuming, we face two problems of fine-tuning and cosmic coincidence.\\
~~Thus, dynamical dark energy models become popular, what those models that originate from various fields, what those models that probe the nature of DE, according to some basic quantum gravitational principles. One example of latter paradigm is the agegraphic DE (ADE) model which has originated from quantum gravity and posseses some of its significant features. In this model, the age of the Universe $T=\int dt$ is used as the IR cut-off L\cite{Cai2007}. But this scenario cannot justify the matter-dominated era, thus it was generalized to the agegraphic dark energy, namely the use of the conformal time $\eta$ as the IR cutoff L \cite{Wei,Wu}.\\ 
On the other hand, a theory of gravity renormalizable with higher spatial derivatives in four dimensions which is also similar to a scalar field theory of Lifshitz \cite{Lifshitz}, in which the time dimension has weight 3 if a space dimension has weight1, almost ten years ago was proposed by Horava \cite{Horava},for this reason, this theory is called Horava-Lifshitz gravity. In the expression of characteristics of this theory can be said that , the causal structures of this theory are different from those present in General Relativity \cite{Greenwald}, because this theory is not Lorentz invariant and since it is non-relativistic, consequently, the speed of light $c$ diverges in the Ultra-Violet (UV) limit.
 For a recent review on Horava-Lifshitz (HL) gravity \cite{Chen, Chen1, Chen2} and it\textquoteright s application as the cosmological framework of the Universe see \cite{Calcagni, Calcagni1, Calcagni2, Calcagni3, Calcagni4, Calcagni5, Calcagni6,Calcagni7,Calcagni8, Calcagni9, Calcagni10, Calcagni11}. 
The logarithmic new agegraphic dark energy model has been studied in the framework of Horava-Lifshitz cosmology \cite{star1}.
 The generalized second law of thermodynamics and new agegraphic dark energy model have also been checked in Horava-Lifshitz cosmology \cite{star3, star4}. Solving the flatness problem with an anisotropic instanton in HL gravity has been studied in \cite{star5}. In \cite{star6} it is analyzed the electromagnetic- gravity interaction in HL framework. A noncommutative version of Friedmann Robertson -Walker (FRW) cosmological models in HL theory has been studied in \cite{star7} as well as the Hamiltonian dynamics of bouncing Bianchi IX cosmologies is examined \cite{star8}. The phase space analysis of Horava- Lifshitz cosmology for a wide range of self- interacting potentials has been studied in \cite{star9} as well as bouncing cosmology for entropy corrected models in
Horava-Lifshitz gravity has also been checked in \cite{star10}.\\  Similarly to the black hole, we use holographic principle in the cosmological applications, because of the fact that the entropy of the Universe can be counted by assuming that the universe is seen as a two-dimensional structure on the horizon of cosmology. Gibbs in his book with title \textquotedblleft Elementary Principles in Statistical Mechanics\textquotedblright \cite{Gibbs} pointed out that systems have a long range interaction, such as gravitation, do not necessarily obey the Boltzmann-Gibbs (BG) theory, and indeed these systems can violate the extensivity constraint of the Boltzmann-Gibbs entropy. On this basis, Tsallis in 1988 \cite{Tsallis}, introduced a non-additive entropy for the non-extensive systems which can be written in compact form as \cite{Tsallis1}
\begin{eqnarray}\label{HL1}
S_{T}=\gamma A^{\delta},
\end{eqnarray}
where $\gamma$ is an unknown constant and $\delta$ denotes the non-extensive parameter. Recently, using relation (\ref{HL1}) and holographic hypothesis, led to the suggestion of a dark energy density in the form 
\begin{eqnarray}\label{Trho}
\rho_D=BL^{2\delta-4},
\end{eqnarray}
where B is an unknown parameter, and it attracts more attemps to itself  \cite{Tavayef2,Tavayef3,Tavayef4,Tavayef5,Tavayef6,Tavayef8,Tavayef9,Tavayef10,Tavayef11,Tavayef12,Tavayef13,Tavayef15,Tavayef16,Tavayef17,Tavayef18,Tavayef19,Tavayef20,Tavayef21,Tavayef22,Tavayef23,Tavayef24,Tavayef25, Hooman25}. Here, we are interested in studying some cosmological consequences by considering the conformal time as the IR cut off in background Horava-Lifshitz cosmology.\\   The paper is organized as follows. In section II, we present Horava-Lifshitz cosmology and we analyze the new Tsallis agegraphic dark energy in Horava-Lifshitz cosmology, both in the simple and in the interacting forms. Beside extracting the differential equation that determines the evolution of the dark energy density parameter, as well as we investigate the model by using the statefinder diagnostic and $\omega-{\omega}^{\prime}$ analysis in section III. The conclusions are given in section IV.

\section{Horava-Lifshitz cosmology}
We begin with a brief review of the cosmological model based on Horava-Lifshitz gravity \cite{Kiritsis}. What the projectability condition dictates is that the lapse function $N$ should be space-indepemdent, while the shift vector  $N^i$ and the 3- dimentional metric $g_{ij}$ aredependent on both time and space. In terms these dynamical variables, the ffull metric is parametrized as
\begin{eqnarray}\label{HL2}
ds^2 = - N^2 dt^2 + g_{ij} (dx^i + N^i dt ) ( dx^j + N^j dt ) ,
\end{eqnarray}


\subsection{Detailed balance}
The gravitational action of HL gravity based on a kinetic and a potential part can be expressed as $S_g = \int dt d^3x \sqrt{g} N ({\cal L}_K+{\cal
L}_V)$ and under the assumption of detailed balance \cite{Horava1}, the full action of HL gravity can be written as follow
\begin{equation}\label{acct}
 S_g =  \int dt d^3x \sqrt{g} N \left\{
\frac{2}{\kappa^2} (K_{ij}K^{ij} - \lambda K^2) +\frac{\kappa^2}{2
w^4} C_{ij}C^{ij}+L1+L2\right\}
\end{equation}
\begin{equation}\nonumber
 L1=-\frac{\kappa^2 \mu}{2 w^2}
\frac{\epsilon^{ijk}}{\sqrt{g}} R_{il} \nabla_j R^l_k +
\frac{\kappa^2 \mu^2}{8} R_{ij} R^{ij}
\end{equation}
\begin{equation}\nonumber
 L2=+\frac{\kappa^2 \mu^2}{8( 3 \lambda-1)} \left[ \frac{1 -
4 \lambda}{4} R^2 + \Lambda  R - 3 \Lambda ^2 \right] ,
\end{equation}
where $K_{ij}$ is the extrinsic curvature which takes the form 
\begin{eqnarray}
K_{ij} = \frac{1}{2N} \left( {\dot{g}_{ij}} - \nabla_i N_j -
\nabla_j N_i \right)
\end{eqnarray}
and a dot denotes a derivative with respect to $t$ and covariant derivatives defined with respect  to the spatial metric $g_{ij}$.
Also 
\begin{eqnarray} C^{ij} \, = \, \frac{\epsilon^{ijk}}{\sqrt{g}} \nabla_k
\bigl( R^j_i - \frac{1}{4} R \delta^j_i \bigr)
\end{eqnarray}
is the Cotton tensor, $\lambda$, $\epsilon^{ijk}$ and $\Lambda $ are a dimensionless constant, the
totally antisymmetric unit tensor and the cosmological constant, respectively. Finally, the  variables $\kappa$, $w$ and $\mu$ are constant parameters with mass dimensions $-1$, $0$ and $1$, respectively.
For focussing on cosmological contents, we should impose the so called projectability condition \cite{Kiritsis} under the detailed balance, then we consider a Friedmann-Robertson-Walker (FRW) metric,
 \begin{eqnarray}
N=1~,~~g_{ij}=a^2(t)\gamma_{ij}~,~~N^i=0~,
\end{eqnarray}
with
\begin{eqnarray}
\gamma_{ij}dx^idx^j=\frac{dr^2}{1- k r^2}+r^2d\Omega_2^2~,
\end{eqnarray}
where $ k=-1,0,+1$ refer to spatially open, flat, and closed
universe respectively. Taking the variation of action with respect to the metric components $N$ and $g_{ij}$, we can obtain the equation of motion as 
\begin{eqnarray}\nonumber
H^2 = \frac{\kappa^2}{6(3\lambda-1)} \rho_m
+\frac{\kappa^2}{6(3\lambda-1)}\left[ \frac{3\kappa^2\mu^2
k^2}{8(3\lambda-1)a^4} +\frac{3\kappa^2\mu^2\Lambda
^2}{8(3\lambda-1)}
 \right]
\end{eqnarray}
\begin{eqnarray}\label{Fr1fluid}
-\frac{\kappa^4\mu^2\Lambda  k}{8(3\lambda-1)^2a^2}
\end{eqnarray}
\begin{eqnarray}\nonumber
\dot{H}+\frac{3}{2}H^2 = -\frac{\kappa^2}{4(3\lambda-1)} p_m
\end{eqnarray}
\begin{eqnarray}\label{Fr2fluid}
-\frac{\kappa^2}{4(3\lambda-1)}\left[\frac{\kappa^2\mu^2
k^2}{8(3\lambda-1)a^4} -\frac{3\kappa^2\mu^2\Lambda
^2}{8(3\lambda-1)}
 \right]-\frac{\kappa^4\mu^2\Lambda  k}{16(3\lambda-1)^2a^2} ,
\end{eqnarray}
where we have defined the Hubble parameter as $H\equiv\frac{\dot
a}{a}$ and $a$ is scale factor. Also $\rho_m$ and $p_m$ are corresponding to energy density and pressure of the matter. At this stage, by noticing the form of the preceding Friedmann equations, the energy density $\rho_D$ and pressure $p_D$ for dark energy can define as   
\begin{equation}\label{rhoDE}
\rho_{D}\equiv \frac{3\kappa^2\mu^2 k^2}{8(3\lambda-1)a^4}
+\frac{3\kappa^2\mu^2\Lambda ^2}{8(3\lambda-1)}
\end{equation}
\begin{equation}\label{pDE} 
p_{D}\equiv \frac{\kappa^2\mu^2
k^2}{8(3\lambda-1)a^4} -\frac{3\kappa^2\mu^2\Lambda
^2}{8(3\lambda-1)}.
\end{equation}
  It is interesting to note that the first term on the right-hand side proportional to $a^{-4}$ is the usual ``dark radiation
term'', present in Ho\v{r}ava-Lifshitz cosmology \cite{Kiritsis}, while the second term is just the explicit cosmological constant.\\   As a last step, these expressions reduce to the standard Friedmann equations $(c=1)$, provided we consider \cite{Kiritsis}
\begin{eqnarray}\label{HL3}
&&G_{\rm cosmo}=\frac{\kappa^2}{16\pi(3\lambda-1)}\\
&&\frac{\kappa^4\mu^2\Lambda}{8(3\lambda-1)^2}=1,
\end{eqnarray}
where $G_{\rm cosmo}$ presents the Newton's cosmological constant. It is worth mentioning that in gravitational theories with the violation of Lorentz invariance, $G_{\rm g}$ in the
gravitational action differs from $G_{\rm cosmo}$ in the Friedmann equations, unless Lorentz invariance is restored \cite{Lim}.
 For completeness we define $G_{\rm g}$ as
\begin{eqnarray}\label{Ggrav}
G_{\rm g}=\frac{\kappa^2}{32\pi}.
\end{eqnarray}
Clearly, in the IR limit ($\lambda=1$), where
Lorentz invariance is restored, $G_{\rm cosmo}$ and $G_{\rm g}$ are the same.\\ Now, by using the above identifications, we can rewrite the modified Friedmann equations (\ref{Fr1fluid})and (\ref{Fr2fluid}) in the standard form as
\begin{equation}\label{Fr1b}
 H^2+\frac{k}{a^2} = \frac{8\pi G_{\rm
cosmo}}{3}\left(\rho_m+\rho_{D}\right)
\end{equation}
\begin{equation}\label{Fr2b}
 \dot{H}+\frac{3}{2}H^2+\frac{k}{2a^2} = - 4\pi G_{\rm
cosmo}\left(p_m+p_D\right).
\end{equation}
\section{ New Tsallis agegraphic dark energy model (NTADE)}
Here we would like to study the NTADE \cite{Abdollahi 2} in HL theory, to do this, we consider a spatially non-flat Universe in which there are a new Tsallis  agegraphic dark energy $\rho_D$ and pressureless dark matter $\rho_m$. If we introduce the density parameters as
\begin{equation}\label{densityparam}
\Omega_m=\frac{8\pi G_{\rm cosmo}}{3H^2}\rho_m,\ \ \
\Omega_{D}=\frac{8\pi G_{\rm cosmo}}{3H^2}\rho_{D},\ \
\Omega_k=\frac{k}{a^2H^2}, 
\end{equation}
then, the first Friedmann equation (\ref{Fr1b}) can be rewritten as
\begin{equation}\label{Fri2}
1+\Omega_k=\Omega_m+\Omega_D.
\end{equation}
As mentioned earlier, because the original ADE model has some difficulties in particular, in to justify the matter-dominated era \cite{Cai}, it motivated Wei and Cai \cite{Wei} to propose the new ADE model, while the time scale is chosen to be the conformal time instead of the age of the Universe. Considering the conformal time as IR cutoff which is difined as $dt=a d\eta$ leading to
$\dot{\eta}=1/a$ and thus
\begin{equation}
\eta=\int_0^a{\frac{da}{Ha^2}}.
\end{equation}
Thus, the corresponding dark energy by considering Eq.(\ref{Trho}) reads
\begin{eqnarray}\label{Tnage}
\rho_D=B{\eta}^{2\delta-4}.
\end{eqnarray}
Taking time derivative of above equation and using $\dot{\eta}=1/a$, we can obtain 
\begin{equation}\label{dotrho}
\dot{\rho}_{D}=\frac{\rho_{D}(2\delta-4)}{a\eta},
\end{equation}
Also, if we take the time derivative of the second relation in (\ref{densityparam}) after using (\ref{dotrho}) and relation $\dot{\Omega}_D={\Omega}^{\prime}_{D}H$, we can obtain the equation of motion for $\Omega_D$ as
\begin{equation}\label{nageOmega}
{\Omega}^{\prime}_{D}=2\Omega_D\left(\frac{\delta-2}{a\eta H}-\frac{\dot{H}}{H^2}\right)
\end{equation}
where prime denotes the derivative with respect to  $x=\ln a$. As previous, by taking derivative of the third relation in (\ref{densityparam}) we get
\begin{equation}
{\Omega}^{\prime}_{k}=-2\Omega_k \left(1+\frac{\dot{H}}{H^2}\right).
\end{equation}

\subsection{statefinder diagnostic and $\omega-{\omega}^{\prime}$ analysis}
However two cosmological parameters $H$ and $q$ are useful to describe the evolution of the Universe, but these two parameters cannot differentiate various dark energy models. For this reason, Sahni et al.,\cite{Sahni} introduced a new geometrical diagnostic pair parameter $\{r,s\}$, knows as the statefinder pair, difined as 
\begin{equation}\label{statefinder}
r=\frac{\dddot{a}}{aH^3},~~~~~~~~~~~~~~ s=\frac{r-1}{3(q-1/2)},
\end{equation}
which clearly show the statefinder pair depend only on the scale factor and it's time derivatives up to third order. Note that the parameter $r$ is also called cosmic jerk and can be expressed in terms of the Hubble and the deceleration parameters as 
\begin{equation}\label{rr2}
r=2q^2+q-\frac{\dot q}{H}.
\end{equation}
  Before we apply the statefinder diagnostic to the NTADE model in HL gravity, it is better to note that i) in the $\{r,s\}$ plane,  $s >0$ ($s<0$) corresponds to a
quintessence-like (phantom-like) model of DE respectively. ii) In a flat $\Lambda$CDM
model and matter dominated universe (SCDM) one finds $\{r,s\}=\{1,0\}$ and $\{r,s\}=\{1,1\}$, respectively. As a complement to statefinder diagnosis, the  $\omega_D-{\omega}^{\prime}_{D}$ analysis is also useful method for distinguish different cosmological models \cite{Caldwell}. In this approach i) the $\Lambda$CDM model corresponds to a fixed point
$\{\omega_D=-1,{\omega}^{\prime}_{D}=0\}$ in the
$\omega_D-{\omega}^{\prime}_{D}$ plane, where ${\omega}^{\prime}_{D}$ represents the derivative of $\omega_D$ with respect to $x=\ln a$. ii) ${\omega}^{\prime}_{D}>0$ and $\omega_D<0$ present the thawing region. (iii) ${\omega}^{\prime}_{D}<0$
and $\omega_D<0$ present the freezing region \cite{Caldwell}.

\subsection{Non-interacting case ($Q=0$)}
As usual, in the new Tsallis agegraphic dark energy scenario, the energy densities for matter and dark energy obey the standard evolution equation:
\begin{eqnarray}\label{conm1}
&&\dot{\rho}_m+3H(\rho_m+p_m)=0,\\
&&\dot{\rho}_D+3H(\rho_D+p_D)=0,\label{conD1}
\end{eqnarray}
where $p_m$ is pressure of matter (here we take $p_m=0$) and $p_D=\omega_D \rho_D$, also, $\omega_D$ is the equation of state (EoS) parameter of the NTADE model what can be achieved by inserting Eq.(\ref{dotrho}) in relation (\ref{conD1}) as
\begin{eqnarray}\label{EoSna}
\omega_D=-1-\frac{2\delta-4}{3a\eta H},
\end{eqnarray}
where $\eta=(\frac{3 H^2 \Omega_D}{8\pi G_{\rm cosmo}B})^{\frac{1}{2\delta-4}}$. Also, the time derivative of Eq.(\ref{EoSna}) with respect to $x=\ln a$, we get
\begin{equation}\label{wage1}
{\omega}^{\prime}_{D}=\frac{(\delta-2)\left(2+2(\delta-2)\Omega_D+a \eta H (-1+3\Omega_D-\Omega_k)\right)}{3 a^2 \eta^2 H^2}.
\end{equation}
Taking the derivative of both side of the Friedmann equation (\ref{Fr1b}) with respect to the cosmic time $t$, and using Eqs.(\ref{conm1}) and (\ref{dotrho}) we find
\begin{eqnarray}\label{agedotH}
\frac{\dot{H}}{H^2}=\Omega_k+\Omega_D \left(-\frac{3}{2}u+\frac{\delta-2}{a\eta
H}\right),
\end{eqnarray}
where $u=\frac{\rho_m}{\rho_D}=-1+\frac{1+\Omega_k}{\Omega_D}$ is the ratio of the energy densities. Finally, by substituting the above equation in relation
\begin{equation}\label{qage}
q\equiv-1-\frac{\dot{H}}{H^2},
\end{equation}
we obtain the expression for the deceleration parameter as 
\begin{equation}\label{qage1}
q=-1-\Omega_k+\Omega_D \left(\frac{3}{2}u+\frac{2-\delta}{a \eta
H}\right).
\end{equation}
\begin{figure}[htp]
\begin{center}
\includegraphics[width=8cm]{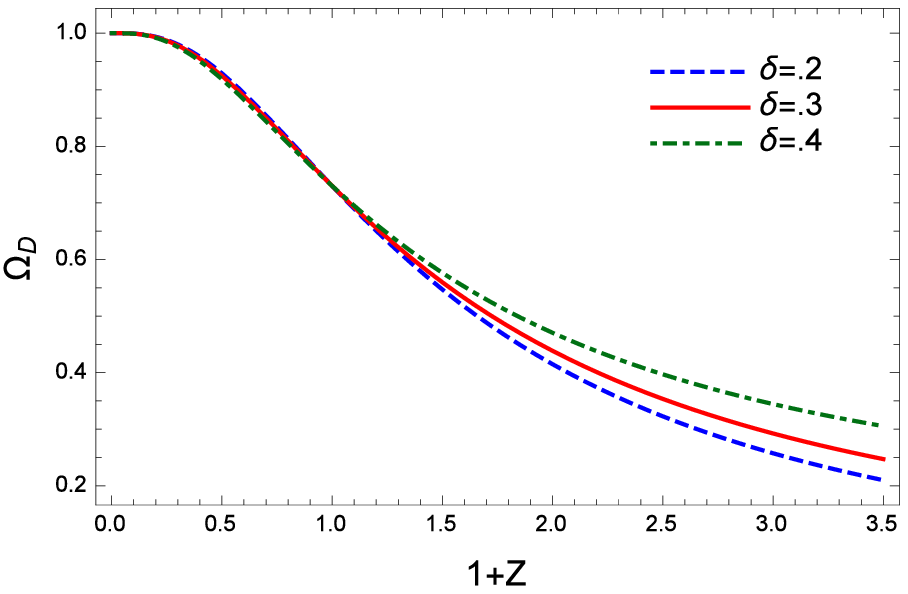}
\caption{The evolution of $\Omega_D$ versus redshift parameter $z$ for
 non-interacting NTADE in HL cosmology. Here, we have taken
$\Omega_D(z=0)=0.73$, $H(z=0)=67$, $\Omega_k(z=0)=0.01$, $\lambda=1.6$ and $B=2.4$ \cite{nonflat}
}\label{Omega-z1}
\end{center}
\end{figure}

\begin{figure}[htp]
\begin{center}
\includegraphics[width=8cm]{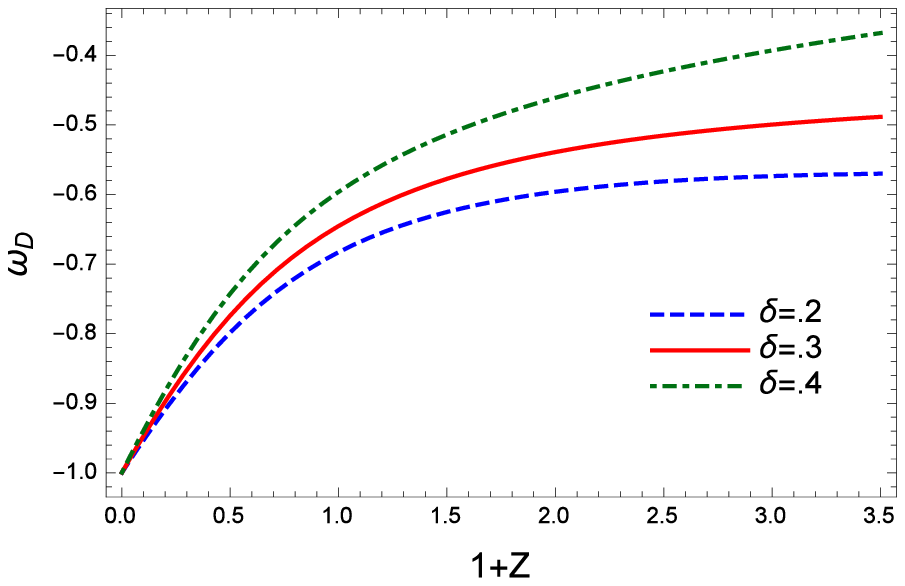}
\caption{The evolution of $\omega_D$ versus redshift parameter $z$ for
 non-interacting NTADE in HL cosmology. Here, we have taken
$\Omega_D(z=0)=0.73$, $H(z=0)=67$, $\Omega_k(z=0)=0.01$, $\lambda=1.6$ and $B=2.4$}\label{w-z1}
\end{center}
\end{figure}

\begin{figure}[htp]
\begin{center}
\includegraphics[width=8cm]{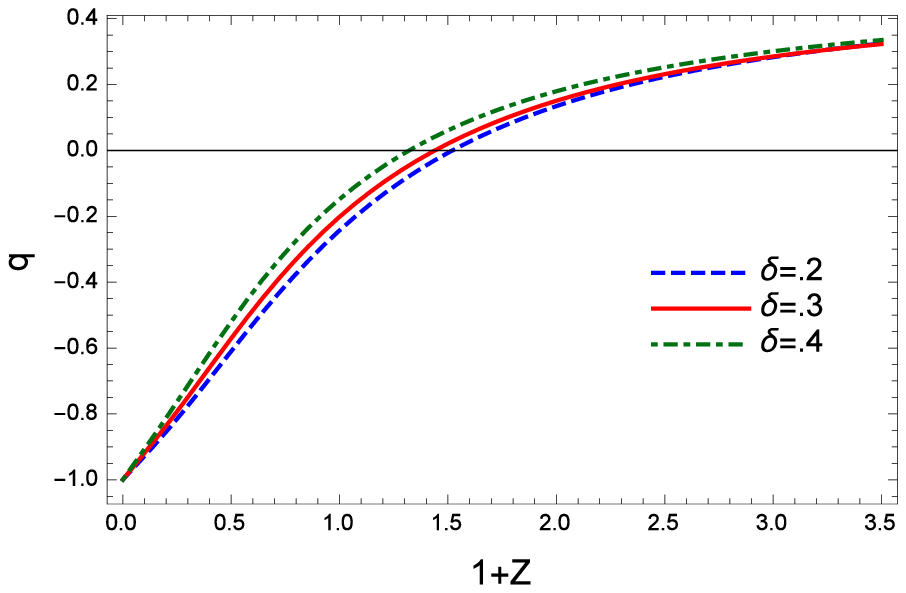}
\caption{The evolution of the deceleration parameter $q$ versus redshift parameter $z$ for
 non-interacting NTADE in HL cosmology. Here, we have taken
$\Omega_D(z=0)=0.73$, $H(z=0)=67$, $\Omega_k(z=0)=0.01$, $\lambda=1.6$ and $B=2.4$}\label{q-z1}
\end{center}
\end{figure}

\begin{figure}[htp]
\begin{center}
\includegraphics[width=8cm]{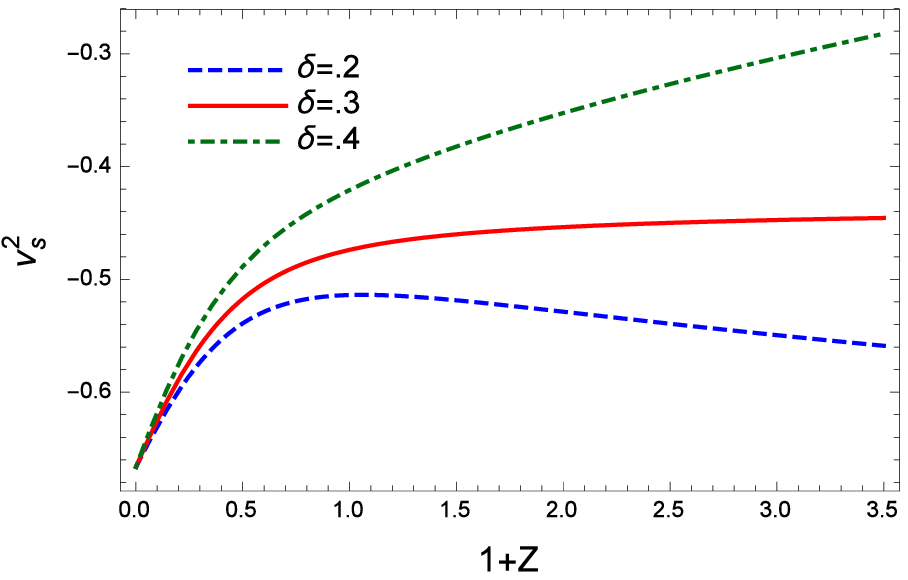}
\caption{The evolution of  the squared of sound speed $v_s^2 $ versus redshift parameter $z$ for
 non-interacting NTADE in HL cosmology. Here, we have taken
$\Omega_D(z=0)=0.73$, $H(z=0)=67$, $\Omega_k(z=0)=0.01$, $\lambda=1.6$ and $B=2.4$}\label{v-z1}
\end{center}
\end{figure}

\begin{figure}[htp]
\begin{center}
\includegraphics[width=6cm]{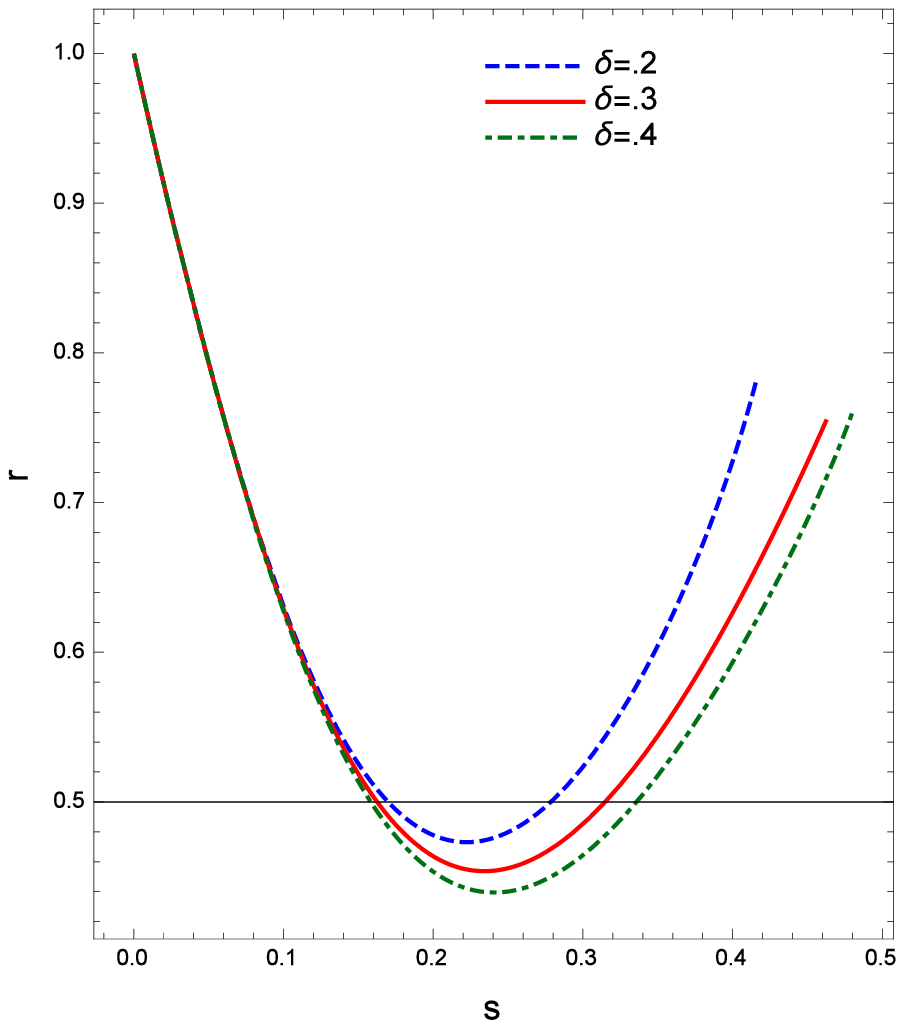}
\caption{The evolution of the statefinder parameter $r$ versus $s$ for
 non-interacting NTADE in HL cosmology. Here, we have taken
$\Omega_D(z=0)=0.73$, $H(z=0)=67$, $\Omega_k(z=0)=0.01$, $\lambda=1.6$ and $B=2.4$}\label{rs-z1}
\end{center}
\end{figure}

\begin{figure}[htp]
\begin{center}
\includegraphics[width=6cm]{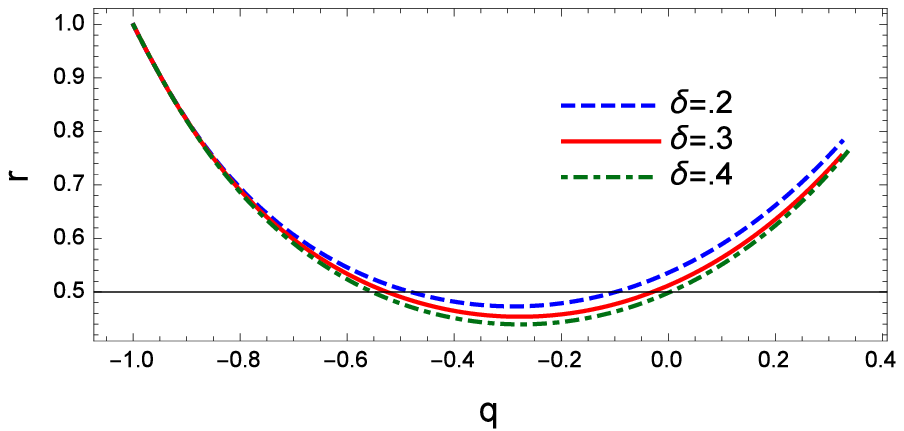}
\caption{The evolution of the statefinder parameter $r$ versus the
deceleration parameter $q$ for non-interacting NTADE in HL cosmology. Here, we have taken
$\Omega_D(z=0)=0.73$, $H(z=0)=67$, $\Omega_k(z=0)=0.01$, $\lambda=1.6$ and $B=2.4$}\label{rq-z1}
\end{center}
\end{figure}

\begin{figure}[htp]
\begin{center}
\includegraphics[width=8cm]{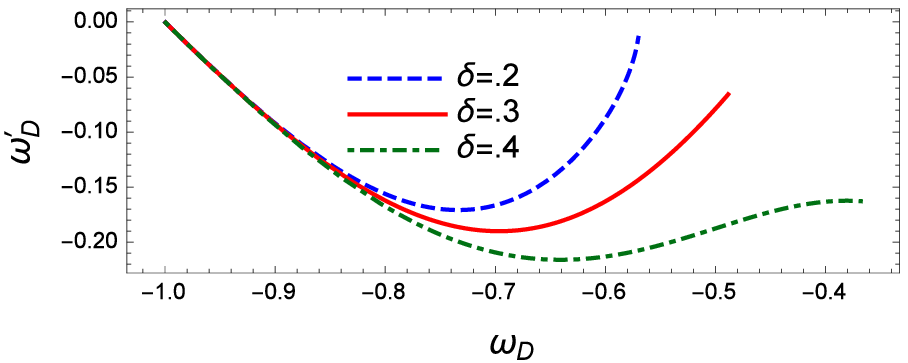}
\caption{The $\omega_D-{\omega}^{\prime}_{D}$ diagram for
non-interacting NTADE in HL cosmology. Here, we have taken
$\Omega_D(z=0)=0.73$, $H(z=0)=67$, $\Omega_k(z=0)=0.01$, $\lambda=1.6$ and $B=2.4$}\label{ww-z1}
\end{center}
\end{figure}

We now consider an important quantity to check the effects of perturbations on the classical stability of our model, namely the square of the sound speed $v_{s}^{2}$, difined as
\begin{equation}\label{vs}
v_{s}^{2}=\frac{dP_D}{d\rho_D}=\frac{\dot{P}_D}{\dot{\rho}_D}=\dfrac{\rho_{D}}{\dot{\rho}_{D}}
\dot{\omega}_{D}+\omega_{D},
\end{equation}
which finally leads to
\begin{equation}\label{Stnage1}
v_{s}^{2}=\frac{5-2\delta+(\delta-2)\Omega_D}{3\eta a H}+\frac{-7+3\Omega_D-\Omega_k}{6},
\end{equation}
for the non-interacting case. The statefinder parameters for NTADE in HL are obtained in accordance with the equation (\ref{statefinder}) as
\begin{equation}\nonumber
r=1+\Omega_k+\frac{(2-\delta)\Omega_D(5-2\delta+(\delta-2)\Omega_D)}{\eta^2 a^2 H^2}
\end{equation}
\begin{equation}\label{rage1}
+\frac{(2-\delta)\Omega_D(-7+3\Omega_D-\Omega_k)}{2\eta a H},
\end{equation}

\begin{equation}\nonumber
s=\frac{(\delta-2)\Omega_D \left(a \eta H (-7+3\Omega_D)+2(5-2\delta+(\delta-2)\Omega_D)\right)}{3a\eta H \left((-4+2\delta+3 a \eta H)\Omega_D-a \eta H \Omega_k \right)}
\end{equation}
\begin{equation}\label{sage1}
-\frac{(2 a \eta H+(\delta-2)\Omega_D)\Omega_k}{3 \left((-4+2\delta+3 a \eta H)\Omega_D-a \eta H \Omega_k \right)}.
\end{equation}
  The evolution of the system parameters for non-interacting case are plotted in
Figs.~\ref{Omega-z1}-\ref{ww-z1}. In Figs.~\ref{Omega-z1}-\ref{q-z1}, we plot the evolution of $\Omega_D$, $\omega_D$ and $q$ versus redshift parameter $z$ for non-interacting case. it is obvious that $\Omega_D$ tends to $0$
in the early universe, $\omega_D$ cannot cross phantom line and the universe enter the acceleration phase earlier for smaller $\delta$. In Fig.~\ref{v-z1} we show the evolution of ${v}^{2}_{s}$, which is unstable here. In Figs.~\ref{rs-z1}-\ref{ww-z1}, we also plot the trajectories of statefinder pair and $\omega-{\omega}^{\prime}$ plane. From trajectory of $(r-s)$, we see a quintessence-ilke behaviour as well as $\omega-{\omega}^{\prime}$ plane presents the freezing region.

\subsection{Interacting case ($Q\neq0$)}
\begin{figure}[htp]
\begin{center}
\includegraphics[width=8cm]{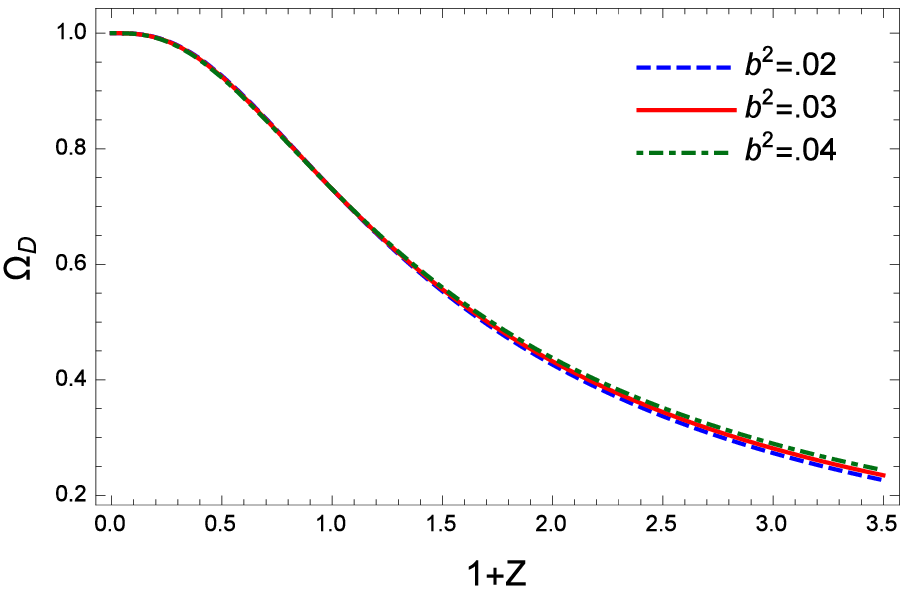}
\caption{The evolution of $\Omega_D$ versus redshift parameter $z$ for
 interacting NTADE in HL cosmology. Here, we have taken
$\Omega_D(z=0)=0.73$, $H(z=0)=67$, $\Omega_k(z=0)=0.01$, $B=2.4$, $\lambda=1.6$ and $\delta=.2$
}\label{Omega-z2}
\end{center}
\end{figure}

\begin{figure}[htp]
\begin{center}
\includegraphics[width=8cm]{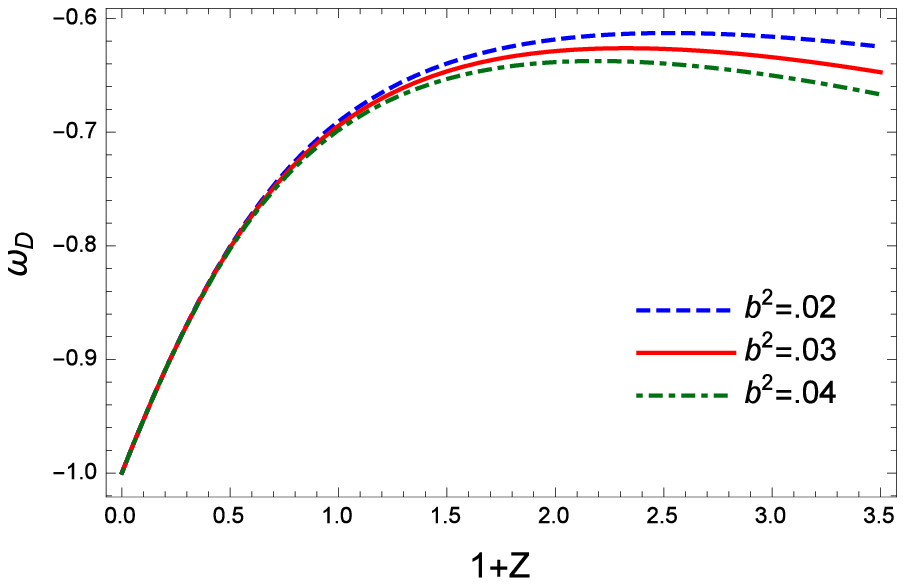}
\caption{The evolution of $\omega_D$ versus redshift parameter $z$ for
 interacting NTADE in HL cosmology. Here, we have taken
$\Omega_D(z=0)=0.73$, $H(z=0)=67$, $\Omega_k(z=0)=0.01$, $B=2.4$, $\lambda=1.6$ and $\delta=.2$}\label{w-z2}
\end{center}
\end{figure}

\begin{figure}[htp]
\begin{center}
\includegraphics[width=8cm]{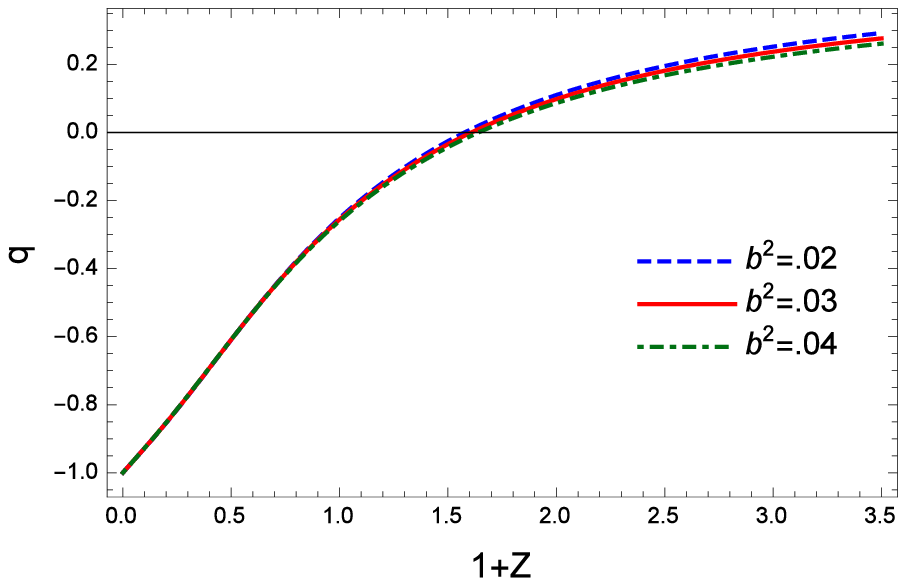}
\caption{The evolution of the deceleration parameter $q$ versus redshift parameter $z$ for
 interacting NTADE in HL cosmology. Here, we have taken
$\Omega_D(z=0)=0.73$, $H(z=0)=67$, $\Omega_k(z=0)=0.01$, $B=2.4$, $\lambda=1.6$ and $\delta=.2$}\label{q-z2}
\end{center}
\end{figure}

\begin{figure}[htp]
\begin{center}
\includegraphics[width=8cm]{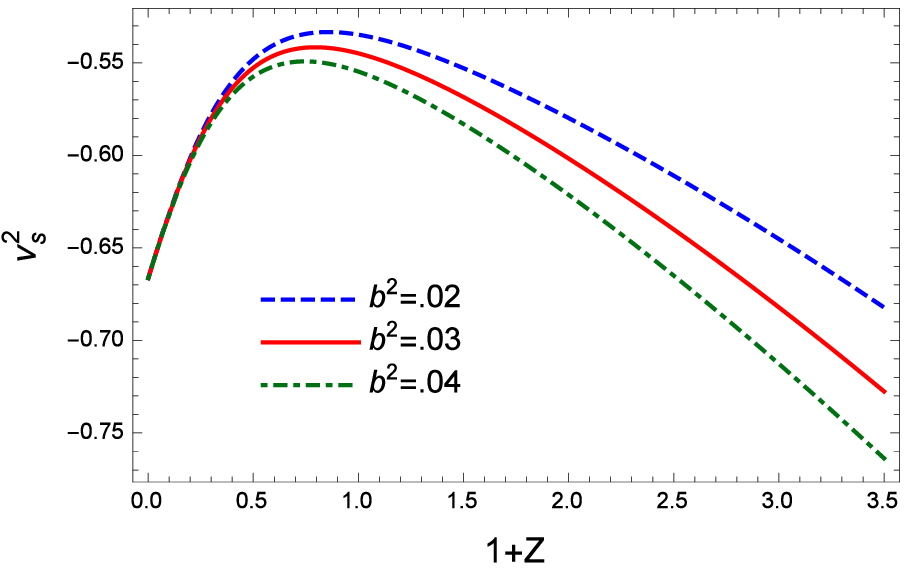}
\caption{The evolution of  the squared of sound speed $v_s^2 $ versus redshift parameter $z$ for
 interacting NTADE in HL cosmology. Here, we have taken
$\Omega_D(z=0)=0.73$, $H(z=0)=67$, $\Omega_k(z=0)=0.01$, $B=2.4$, $\lambda=1.6$ and $\delta=.2$}\label{v-z2}
\end{center}
\end{figure}

\begin{figure}[htp]
\begin{center}
\includegraphics[width=6cm]{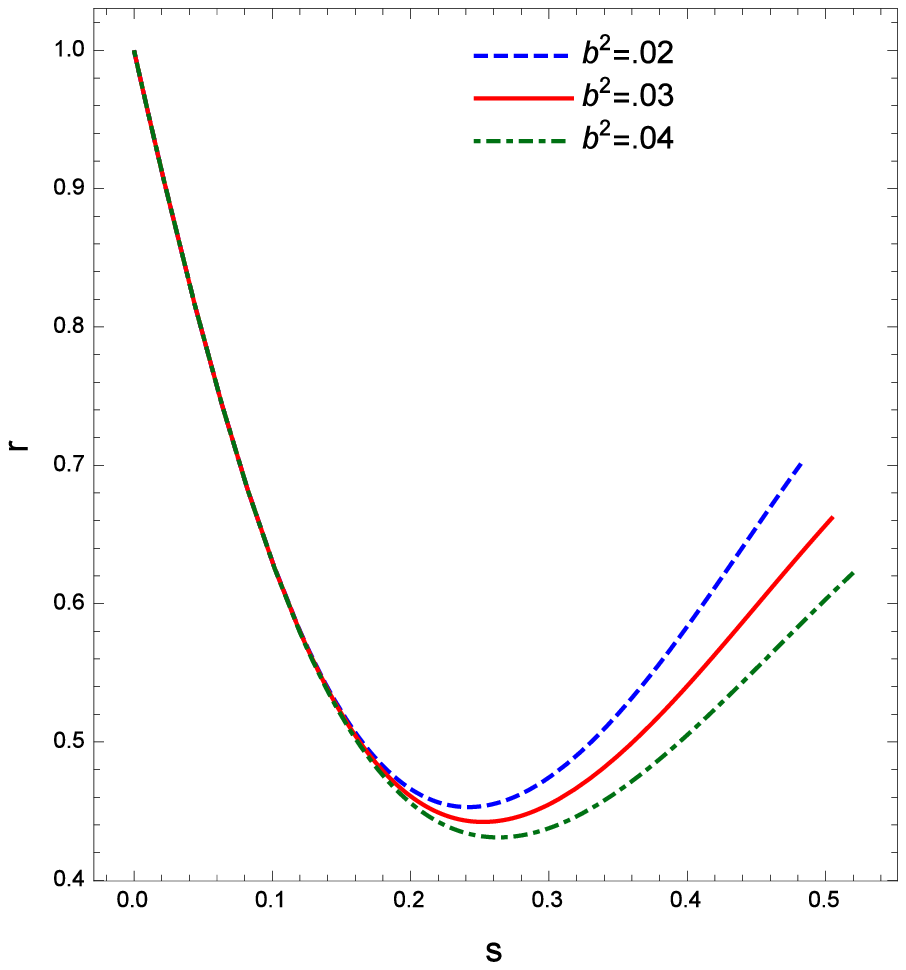}
\caption{The evolution of the statefinder parameter $r$ versus $s$ for
 interacting NTADE in HL cosmology. Here, we have taken
$\Omega_D(z=0)=0.73$, $H(z=0)=67$, $\Omega_k(z=0)=0.01$, $B=2.4$, $\lambda=1.6$ and $\delta=.2$}\label{rs-z2}
\end{center}
\end{figure}

\begin{figure}[htp]
\begin{center}
\includegraphics[width=6cm]{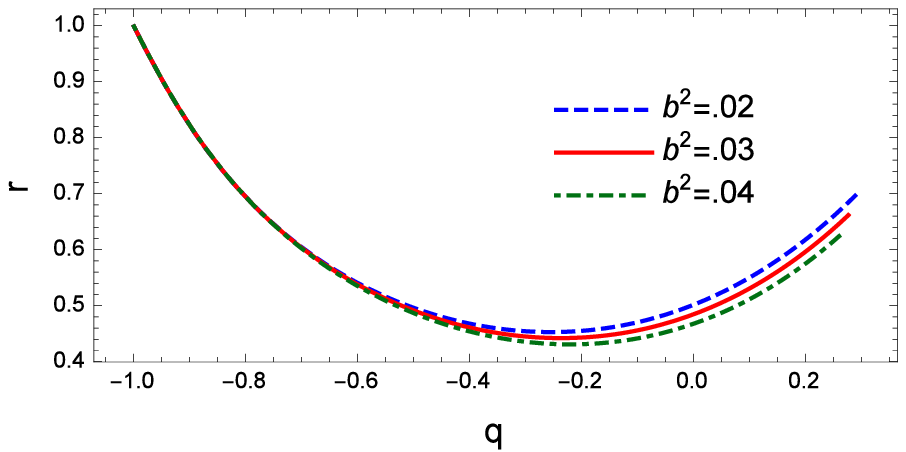}
\caption{The evolution of the statefinder parameter $r$ versus the
deceleration parameter $q$ for interacting NTADE in HL cosmology. Here, we have taken
$\Omega_D(z=0)=0.73$, $H(z=0)=67$, $\Omega_k(z=0)=0.01$, $B=2.4$, $\lambda=1.6$ and $\delta=.2$}\label{rq-z2}
\end{center}
\end{figure}

\begin{figure}[htp]
\begin{center}
\includegraphics[width=8cm]{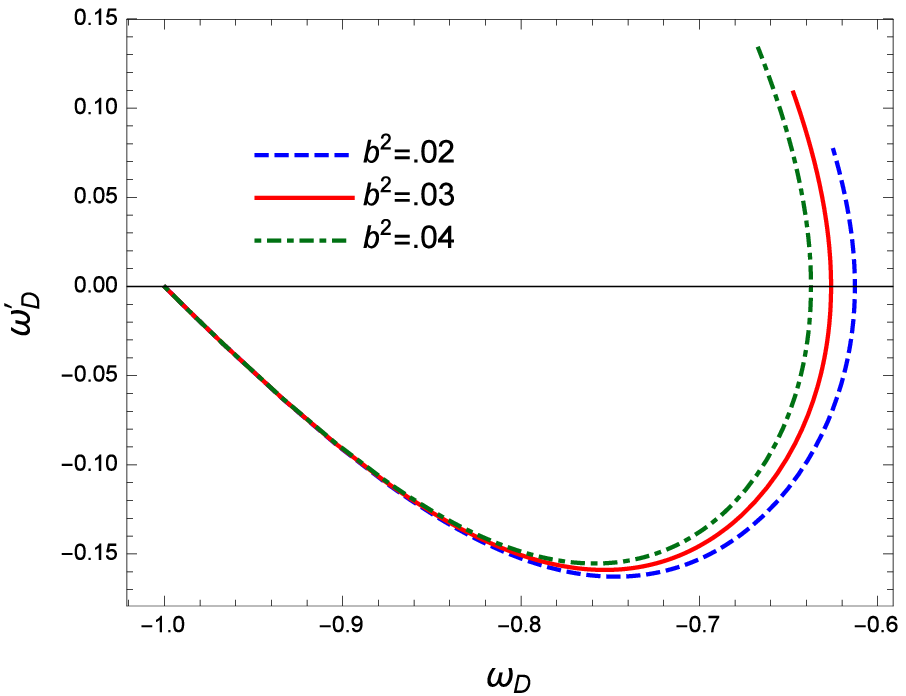}
\caption{The $\omega_D-{\omega}^{\prime}_{D}$ diagram for
 interacting NTADE in HL cosmology. Here, we have taken
$\Omega_D(z=0)=0.73$, $H(z=0)=67$, $\Omega_k(z=0)=0.01$, $B=2.4$, $\lambda=1.6$ and $\delta=.2$}\label{ww-z2}
\end{center}
\end{figure}

Here, we postulate that the two sectors the NTADE and dark matter (DM) interact through the interaction term $Q$, since such a scenario could alleviate the known coincidence problem \cite{Steinhardt1}. It should be noted that the recent observational evidence by the galaxy clusters supports the interaction between DE and DM \cite{Bertolami}. This causes the energy conservaition law for each dark component to be hold separately i.e.
\begin{eqnarray}\label{conm}
&&\dot{\rho}_m+3H\rho_m=Q,\\
&&\dot{\rho}_D+3H(1+\omega_D)\rho_D=-Q,\label{conD}
\end{eqnarray}
where $Q$ has a form as follows $Q=3b^2 H \rho_m$ with $b^2$ being a coupling constant. It deserves mention three cases a bout interaction term $Q$ i) for $Q>0$, there is an energy transfer from NTADE to DM. ii) the form of $Q$ is chosen purely phenomenologically, in order to obtain desirable cosmological results including phantom crossing and accelerated expansion. iii) easily, one can find numerous form of $Q(H\rho)$ in Ref\cite{Jamil, int1, int2, int3, int4}.\\   By repeating the above procedure in the case where the matter and dark sectors are allowed to interact, we find expressions for cosmological application , similary to the previous subsection, 

\begin{eqnarray}\label{NagedotH}
\frac{\dot{H}}{H^2}=\Omega_k+\Omega_D \left(\frac{3}{2}u(b^2-1)+\frac{\delta-2}{a\eta
H}\right),
\end{eqnarray}

\begin{eqnarray}\label{EoSnna}
\omega_D=-1-b^2 u-\frac{2\delta-4}{3a\eta H},
\end{eqnarray}

\begin{equation}\label{qnage1}
q=-1-\Omega_k-\Omega_D \left(\frac{3}{2}u(b^2-1)+\frac{\delta-2}{a\eta
H}\right),
\end{equation}

\begin{equation}\nonumber
v_{s}^{2}=\frac{5-2\delta+(\delta-2)\Omega_D}{3\eta a H}
\end{equation}
\begin{equation}\nonumber
+\frac{-3b^2(\Omega_D-1-\Omega_k)-7+3\Omega_D-\Omega_k}{6}
\end{equation}
\begin{equation}
+\frac{3b^2(b^2-1)a \eta H (\Omega_D-1-\Omega_k)}{2(\delta-2)\Omega_D},
\end{equation}\label{Stnage2}

\begin{equation}\nonumber
r=\frac{(2-\delta)\Omega_D(5-2\delta+(\delta-2)\Omega_D)}{\eta^2 a^2 H^2}
\end{equation}
\begin{equation}\nonumber
+\frac{(2-\delta)\Omega_D(-7+3\Omega_D-\Omega_k+3b^2(1-\Omega_D+\Omega_k))}{2\eta a H}
\end{equation}
\begin{equation}
+\frac{2(1+\Omega_k)+9b^2(b^2-1)(1-\Omega_D+\Omega_k)}{2},
\end{equation}\label{rage2}
\begin{equation}
s=\frac{M1+M2}{M3}.
\end{equation}
\begin{equation}\nonumber
M1=-2(\delta-2)\Omega_D\left(5-2\delta+(\delta-2)\Omega_D\right)
\end{equation}
\begin{equation}\nonumber
+a\eta H \Omega_D(\delta-2)\left(7-3b^2+3(b^2-1)\Omega_D+\Omega_k-3b^2\Omega_k\right)\nonumber\\
\end{equation}
\begin{equation}\nonumber
M2=a^2 {\eta}^2 H^2\left(-9b^2(b^2-1)(\Omega_D-1)+(2+9b^2(b^2-1))\Omega_k\right)\nonumber\\
\end{equation}
\begin{equation}\nonumber
M3=3a\eta H\left[-2(\delta-2)\Omega_D+a\eta H\left(-3b^2+3\Omega_D(b^2-1)+m3\right)\right]\nonumber\\
\end{equation}
\begin{equation}\nonumber
m3=+\Omega_k-3b^2\Omega_k
\end{equation}
In Figs.~\ref{Omega-z2}-\ref{ww-z2}, we present system parameters for various values of $b^2$ for interacting case.

\begin{figure}[htp]
\begin{center}
\includegraphics[width=8cm]{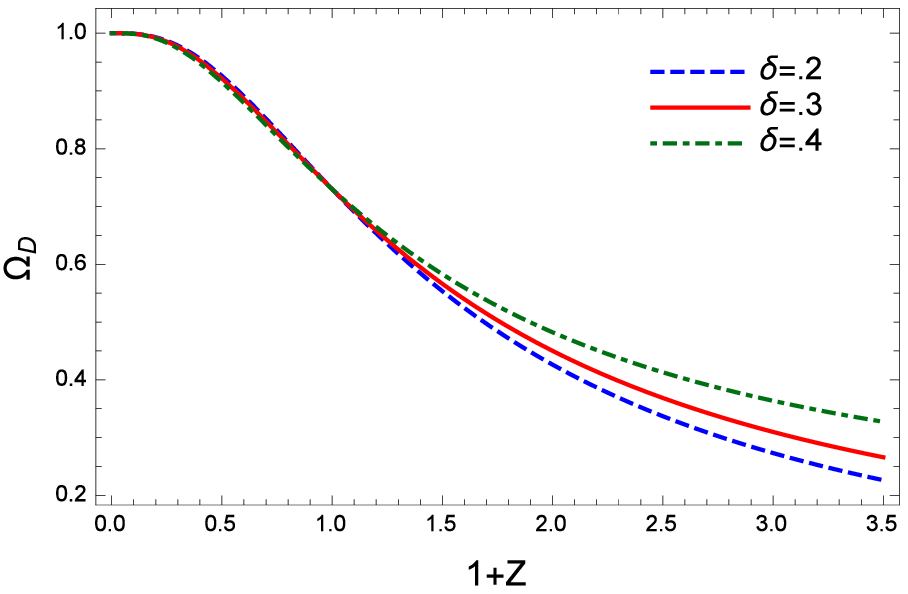}
\caption{The evolution of $\Omega_D$ versus redshift parameter $z$ for
 interacting NTADE in HL cosmology. Here, we have taken
$\Omega_D(z=0)=0.73$, $H(z=0)=67$, $\Omega_k(z=0)=0.01$, $B=2.4$, $\lambda=1.6$ and $b^2=.02$
}\label{Omega-z3}
\end{center}
\end{figure}

\begin{figure}[htp]
\begin{center}
\includegraphics[width=8cm]{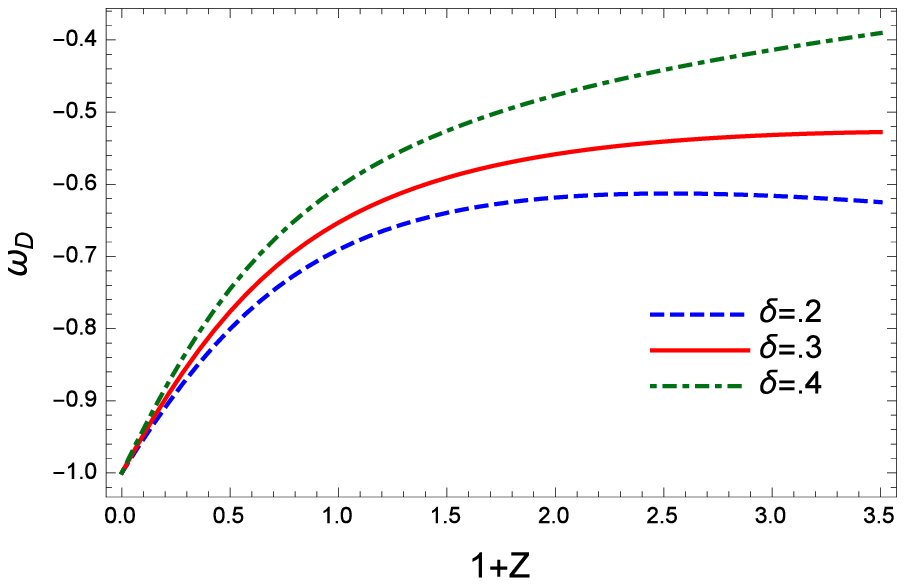}
\caption{The evolution of $\omega_D$ versus redshift parameter $z$ for
 interacting NTADE in HL cosmology. Here, we have taken
$\Omega_D(z=0)=0.73$, $H(z=0)=67$, $\Omega_k(z=0)=0.01$, $B=2.4$, $\lambda=1.6$ and $b^2=.02$}\label{w-z3}
\end{center}
\end{figure}

\begin{figure}[htp]
\begin{center}
\includegraphics[width=8cm]{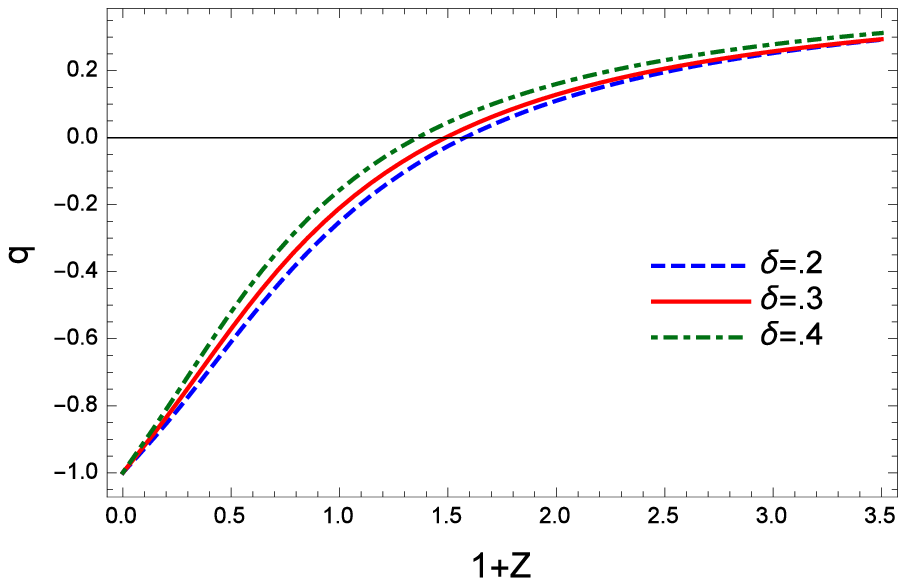}
\caption{The evolution of the deceleration parameter $q$ versus redshift parameter $z$ for
 interacting NTADE in HL cosmology. Here, we have taken
$\Omega_D(z=0)=0.73$, $H(z=0)=67$, $\Omega_k(z=0)=0.01$, $B=2.4$, $\lambda=1.6$ and $b^2=.02$}\label{q-z3}
\end{center}
\end{figure}

\begin{figure}[htp]
\begin{center}
\includegraphics[width=8cm]{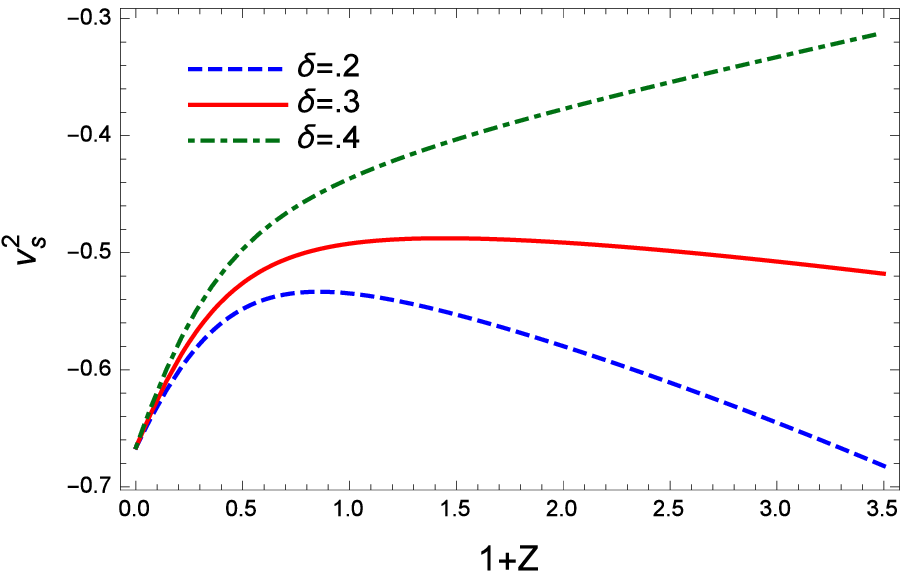}
\caption{The evolution of  the squared of sound speed $v_s^2 $ versus redshift parameter $z$ for
 interacting NTADE in HL cosmology. Here, we have taken
$\Omega_D(z=0)=0.73$, $H(z=0)=67$, $\Omega_k(z=0)=0.01$, $B=2.4$, $\lambda=1.6$ and $b^2=.02$}\label{v-z3}
\end{center}
\end{figure}

\begin{figure}[htp]
\begin{center}
\includegraphics[width=6cm]{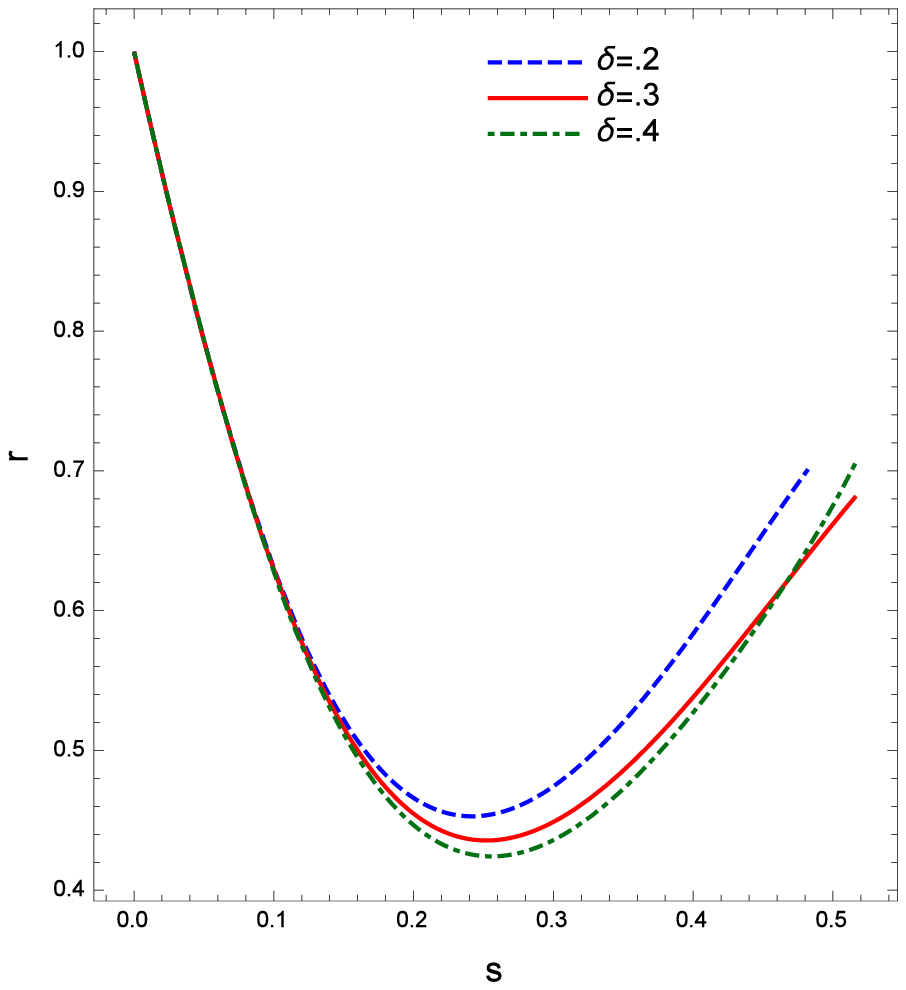}
\caption{The evolution of the statefinder parameter $r$ versus $s$ for
 interacting NTADE in HL cosmology. Here, we have taken
$\Omega_D(z=0)=0.73$, $H(z=0)=67$, $\Omega_k(z=0)=0.01$, $B=2.4$, $\lambda=1.6$ and $b^2=.02$}\label{rs-z3}
\end{center}
\end{figure}

\begin{figure}[htp]
\begin{center}
\includegraphics[width=6cm]{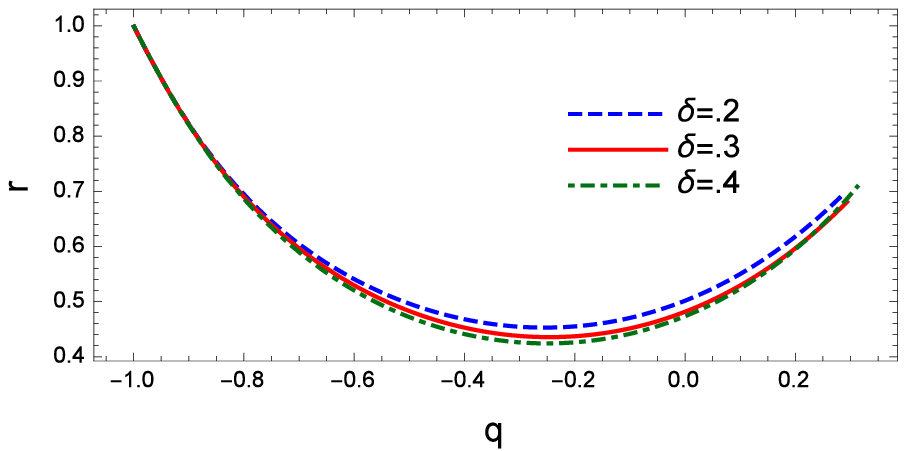}
\caption{The evolution of the statefinder parameter $r$ versus the
deceleration parameter $q$ for interacting NTADE in HL cosmology. Here, we have taken
$\Omega_D(z=0)=0.73$, $H(z=0)=67$, $\Omega_k(z=0)=0.01$, $B=2.4$, $\lambda=1.6$ and $b^2=.02$}\label{rq-z3}
\end{center}
\end{figure}

\begin{figure}[htp]
\begin{center}
\includegraphics[width=8cm]{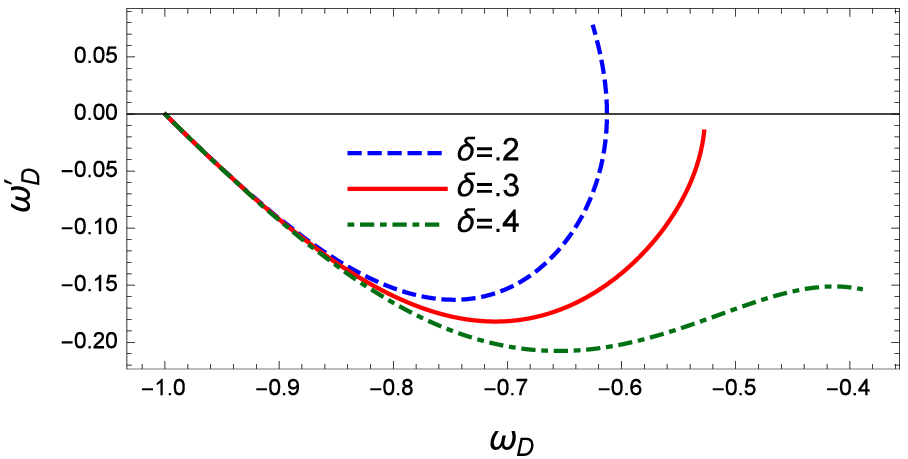}
\caption{The $\omega_D-{\omega}^{\prime}_{D}$ diagram for
 interacting NTADE in HL cosmology. Here, we have taken
$\Omega_D(z=0)=0.73$, $H(z=0)=67$, $\Omega_k(z=0)=0.01$, $B=2.4$, $\lambda=1.6$ and $b^2=.02$}\label{ww-z3}
\end{center}
\end{figure}

In Fig. \ref{w-z2}, we see that the equation of state $\omega_D$ cannot cross phantom line as well as by considering different values of $b^2$, Fig. \ref{v-z2}, shows unstability of this model. 
The evolutionary trajectories for $(r-s)$ and $(r-q)$ planes for TNADE model in HL cosmology have plotted in Figs. \ref{rs-z2} and \ref{rq-z2} respectively. Fig. \ref{rs-z2}, shows that, the evolutionary trajectories $r$ and $s$ end at $\Lambda CDM$ $(r=1,s=0)$ in the future for different values of $b^2$ as well as from Fig. \ref{rq-z2}, we see that the evolutionary trajectories started from matter dominated universe in the past and approach the point $(r=1, q=-1)$ in the future. The $\omega_D-{\omega}^{\prime}_{D}$ plane for TNADE model in HL cosmology, by considering different values of $b^2$ has plotted in Fig. \ref{ww-z2}. We see that this model represents the freezing region. In Figs.~\ref{Omega-z3}-\ref{ww-z3}, we show cosmological parameters for different values of $\delta$ for interacting case. From Figs. \ref{Omega-z3}-\ref{q-z3}, we see that $\Omega_D$ trends to 1 at late time, $\omega_D$ cannot cross phantom line and $q$ shows a transition from the deceleration phase to the acceleration one. In Fig. \ref{v-z3}, we have plotted $v_s^2 $ for different values of $\delta$. We observe that the model is unstable. We have plotted Figs. \ref{rs-z3}-\ref{ww-z3} for statefinder parameters and  $\omega-{\omega}^{\prime}$ plane by considering different values of $\delta$. From Fig. \ref{rs-z3}, we observe that for $(r-s)$ plane of NTADE, the statefinder parameters $r$ and $s$ end at $\Lambda CDM$ $(r=1, s=0)$ in the future by considering interaction term as well as in $(r, q)$ evolutionary plane , we see that the evolutionary trajectories have started from matter dominated universe in the past and end their evolution in $\Lambda CDM$ $(r=1, q=-1)$. In Fig. \ref{ww-z3}, the $\omega-{\omega}^{\prime}$ plane shows the freezing region which correspond to accelerated expansion of the Universe.


\section{Closing remarks}
In this paper, we have investigated the NTADE scenario in the framework of Horava-Lifshitz cosmology. Since ADE density corresponds to a dynamical cosmological constant, we used from a dynamical framework, instead of general relativity. Thus, we investigated the NTADE in the framework of Horava-Lifshitz cosmology.\\   Since experimental data have implied that our universe is not a perfectly flat universe, we imposed an arbitrary curvature for the background geometry, and we allowed for an interacting between the matter and dark energy sectors. For both non-interacting and interacting cases, we extracted the differential equation that determines the evolution of the dark energy density parameter, which gives a suitable estimate for the state parameter of dark energy as well as the deceleration parameter to study an expansion of the universe. Also, we studied statefinder trajectories and $\omega-{\omega}^{\prime}$ plane. To study parametric behaviour, we found that phantom crossing can not occure for the state parameter for small values of coupling parameter $b^2$ and $\delta$, and from the plot of the deceleration parameter, we have observed a transition from decelerating to accelerating phase of the universe. Also, model is not stable while the cosmological plane can meet the freezing region.


\end{document}